\documentclass[5p]{elsarticle}

\journal{Physica E}

\usepackage{color}
\usepackage{amsmath}
\usepackage{amssymb}

\usepackage{color}
\usepackage{amsmath}
\usepackage{graphicx}
\usepackage{multirow}
\usepackage{float}

\DeclareMathOperator{\sgn}{sgn} 

\begin{document}

\newcommand{\bn}{{ n}}
\newcommand{\bp}{{ p}}   
\newcommand{\br}{{ r}}
\newcommand{\bk}{{ k}}
\newcommand{\bv}{{ v}}
\newcommand{\brho}{{\bm{\rho}}}
\newcommand{\bj}{{ j}}
\newcommand{\wk}{\omega_{ k}}
\newcommand{\nk}{n_{ k}}
\newcommand{\eps}{\varepsilon}
\newcommand{\la}{\langle}
\newcommand{\ra}{\rangle}
\newcommand{\be}{\begin{eqnarray}}
\newcommand{\ee}{\end{eqnarray}}
\newcommand{\intl}{\int\limits_{-\infty}^{\infty}}
\newcommand{\dE}{\delta{\cal E}^{ext}}
\newcommand{\SE}{S_{\cal E}^{ext}}
\newcommand{\dsp}{\displaystyle}
\newcommand{\phit}{\varphi_{\tau}}
\newcommand{\p}{\varphi}
\newcommand{\cL}{{\cal L}}
\newcommand{\dphi}{\delta\varphi}
\newcommand{\dbj}{\delta{ j}}
\newcommand{\dI}{\delta I}
\newcommand{\dph}{\delta\varphi}
\newcommand{\tp}{\tilde p}
\newcommand{\uu}{\uparrow\uparrow}
\newcommand{\ud}{\uparrow\downarrow}
\newcommand{\du}{\downarrow\uparrow}
\newcommand{\dd}{\downarrow\downarrow}
\newcommand{\tG}{\tilde\Gamma}
\newcommand{\tP}{\tilde\Phi}
\newcommand{\fe}{\bar f}

\allowdisplaybreaks 

\begin{frontmatter}

\title{Electron-electron scattering and conductance of long many-mode channels}
  
\author{K.~E.~Nagaev}
\address{Kotelnikov Institute of Radioengineering and Electronics, Mokhovaya 11-7, Moscow, 125009 Russia}

\date{\today}

\begin{abstract}
The electron-electron scattering increases the resistance of ballistic many-mode channels whose width
is smaller than their length. We show that this increase saturates in the limit of infinitely long channels. 
Because the mechanisms of angular relaxation of electrons in three and two dimensions are different, 
the saturation value of the correction to the resistance is temperature-independent in the case of 
three-dimensional channels and is proportional to the temperature for two-dimensional ones. The spatial behavior of electron
distribution in the latter case is described by an unusual characteristic length.
\end{abstract}


\end{frontmatter}

\section{Introduction}

Though the electron-electron scattering does not directly contribute to the electrical resistance in
the absence of umklapp processes \cite{Peierls},  it affects the current in small-size  conductors. In
particular, it leads to a minimum in the temperature dependence of the resistance \cite{Molenkamp94} of a wire
with diffusive boundary scattering due to the electronic analogues of Knudsen \cite{Black80} and Poiseuille effects. The latter  represents a decrease of resistance with increasing temperature due to decreasing viscosity of the electron liquid and is also known as the Gurzhi effect \cite{Gurzhi63}. A similar decrease of resistance 
was obtained later for 2D constrictions with viscous electron flow \cite{Guo17}, where the electron-electron 
scattering serves as a "lubricant" for the rough boundaries of the conducting area. The electron-electron 
scattering results in the decrease of the resistance even for contacts with smooth boundaries because it changes 
the trajectories of electrons and may prevent them from passing through the constriction or help them to get 
through it \cite{Nagaev08,Nagaev10}. This decrease was experimentally observed in several 
papers \cite{Renard08,Melnikov12}.

\begin{figure}[t]
\includegraphics[width=5cm]{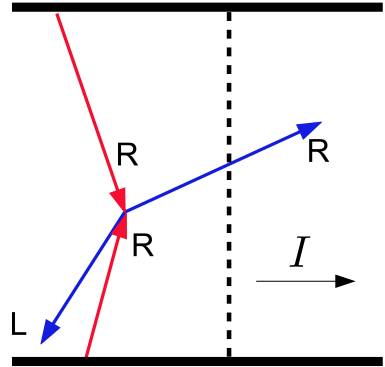}
\caption{\label{coll} A collision of two electrons that changes the number of right-movers. One of
the right-movers is converted into a left-mover despite the momentum conservation.}
\end{figure}

As the electron-electron collisions conserve the total momentum of electrons, they may affect the conductance only 
in the presence of a spatial inhomogeneity that absorbs or provides the extra momentum. In the above cases, this
inhomogeneity was represented by the hard boundaries of the conducting area, but  the extra momentum may be also 
absorbed by the electron reservoirs at the ends of any conducting system of a finite size. This suggests that 
the electron-electron scattering may affect the current in finite-length conducting channels even in the case of 
a specular reflection from the walls. Recently, the correction to the conductance of a narrow multichannel ballistic
conductor was calculated for the weak electron-electron scattering \cite{Nagaev12}. This correction appeared to be
negative and resulted from pairwise collisions that changed the number of electrons moving to the right and to the 
left, i.~e. whose projection of the velocity on the channel axis was positive or negative (see Fig. \ref{coll}). 
In any dimension higher than 1, these collisions are allowed by the conservation laws. If an electron originating 
from one of the reservoirs is scattered back into the same reservoir, it does not contribute to the current and 
hence the resistance of the channel increases \cite{Lunde07}.

As the calculations in Ref. \cite{Nagaev12} were performed in the lowest approximation in the electron-electron scattering, the resulting correction to the conductance was proportional to the length of the channel. 
However, it was not clear whether the conductance tends to zero with increasing length of the channel or stops
to decrease at some finite value. The purpose of the present paper is to calculate the correction
to the conductance in the limit of strong electron-electron scattering. 

The correction to the  electric current is determined by the angular relaxation of electron distribution, which
is essentially different in three-dimensional (3D) and two-dimensional (2D) electron 
gases \cite{Gurzhi95-PRB, Ledwith17}. The 3D relaxation is dominated by small-angle scattering and therefore all
angular harmonics, both odd and even, decay with the same characteristic time. In contrast to this, the 2D relaxation
has a significant contribution from large-angle scattering that results from collisions of electrons with almost
opposite momenta, and this results in strongly different relaxation times of the symmetric and antisymmetric parts of the
distribution function in the momentum space. The angular relaxation of the symmetric part in the 2D case is determined
by the collisions of electrons with almost opposite momenta, which rotate the pair of excess electrons in the momentum space
about the origin, and results in the relaxation rate proportional to $T^2$. In the case of the antisymmetric part, an excess 
electron on one side of the Fermi surface has no pair on its opposite side, and therefore this mechanism does not work. 
Instead the relaxation of this part proceeds through small-angle scattering and its rate is proportional to $T^4$, which 
is much smaller than $T^2$ at low temperatures. Because the odd and even angular harmonics of the electron distribution
are coupled in a spatially inhomogeneous system, determining the temperature dependence of the correction to the 
conductance of the 2D channel is an interesting question. 

The calculation of the correction to the conductance of a long channel presents a nontrivial mathematical problem that cannot
be solved by standard methods of kinetic theory. The first reason is that the calculation of the current involves a large
number of angular harmonics of the electron distribution and not only the lower ones as in bulk conductors.
The second reason is that the electron distribution exhibits a different behavior in different portions of the channel. 
While it is almost constant in its middle part, it sharply changes near its ends, and it is difficult to describe its 
spatial dependence using the same approximations everywhere. To overcome these difficulties, a custom semi-analytical approach is used in this paper.

The paper is organized as follows.  In Sec. II we present the model and basic equations, in Sec. III we perform 
calculations for the 3D case, and Sec. IV presents calculations for the 2D case. In Sec. V we discuss the
results in terms of physics, and Sec. VI presents the summary. Appendices contain more details of calculations.

\section{Model and basic equations}

Consider a metallic wire of a uniform cross-section that connects two electronic reservoirs. We assume that the
length $L$ of the wire is much larger than its transverse dimensions, and these dimensions are much larger 
than the Fermi wavelength. There 
are no impurities in the wire, and the boundaries are assumed to be absolutely smooth so that the electrons are 
specularly reflected from them and their longitudinal momentum is conserved. The narrowness of the channel allows 
us to neglect the effects of electron-electron scattering outside the channel because they are proportional
to the number of transverse quantum modes squared \cite{Nagaev08}.

The distribution function of electrons in the channel obeys the Boltzmann equation 
\be
 \frac{\partial f}{\partial t}
 +
 \bv\,\frac{\partial f}{\partial\br}
 +
 e{ E}\,\frac{\partial f}{\partial\bp}
 =
 \hat{I}_{ee},
 \label{Boltz}
\ee
where ${ E}=-\nabla\phi$ is the electric field and the electron--electron
collision integral $\hat{I}_{ee}$ is given by 
\begin{multline}
 \hat{I}_{ee}(\bp)
 =
 \alpha_{ee}\,\nu_d^{-2}
 \int\frac{d^dk}{(2\pi)^d}
 \int\frac{d^dp'}{(2\pi)^d}
 \int d^dk'\,
\\ {}\times
 \delta(\bp + \bk - \bp' - \bk')\,
 \delta( \eps_{\bp} + \eps_{\bk} - \eps_{\bp'} - \eps_{\bk'} )
\\ {}\times
 \Bigl\{
  [1 - f(\bp)]\,[1 - f(\bk)]\, f(\bp')\, f(\bk')\,
\\{}  -
  f(\bp)\, f(\bk)\, [1 - f(\bp')]\, [1 - f(\bk')]
 \Bigr\},
 \label{I_ee}
\end{multline}
$\alpha_{ee}$ is the dimensionless interaction parameter, $d=2$ or 3 is the dimensionality
of the system; $\nu_3=mp_F/\pi^2$ and $\nu_2=m/\pi$
are the three- and two-dimensional two-spin electronic densities of states ($\hbar=1$).
The assumption of momentum-independent interaction parameter is valid if the screening length of
the electron-electron interaction is sufficiently short. This can be ensured by a high enough
concentration of electrons in the 3D case or by a close electrostatic gate in the 2D case.
The current through an 
arbitrary section of the conductor is given by an integral over the transverse coordinates
\be
 I=
 2e\int d^{d-1}r_{\perp}\,
 \int\frac{d^d p}{(2\pi)^d}\,
 v_x\, f(\bp,x,\br_{\perp}).
 \label{current}
\ee

Because of the condition $E_F \gg{\rm max}(eV, T)$ one may treat the electron velocity near the Fermi surface as 
energy independent and set $\bv = v_F\bn$, where $\bn$ is a unit vector in the direction of $\bp$.
It is possible to avoid solving the Poisson equation for the electric potential $\phi$ if one replaces $\bp$ as the 
argument of $f$ by $\bn$ and the energy variable $\eps = \eps_{\bp} + e\phi(\br) - E_F$. With the new variables, 
the term with electric field drops out from Eq. (\ref{Boltz}), and it takes up the form 
\be
 \frac{\partial f(\bn,\eps,\br)}{\partial t}
 +
 \bv\,\frac{\partial f}{\partial\br}
 =
 \left.\hat{I}_{ee}\{f\}\right|_{\bn,\eps,\br}.
 \label{kinetic}
\ee
The boundary conditions for this equation at the left and right ends of the channel are
\begin{align}
  f(\eps,\, n_x>0,\,x=0) &  = 
  f_0(\eps - eV/2),
  \\
  f(\eps,\, n_x<0,\, x=L) &  = 
  f_0(\eps + eV/2),
 \label{bound_cond}
\end{align}
where $x$ is the longitudinal coordinate, $V$ is the voltage drop across the channel, and 
$f_0(\eps) = 1/[1 + \exp(\eps/T)]$ is the equilibrium Fermi distribution function.

Because we are interested in the electric current, the angular relaxation of electrons will be of
primary importance to us. As the physics of this relaxation  is essentially different in 3D and 2D electron 
gases, one has to make the different approximations for these cases, and in what 
follows we treat them separately.

\section{3D channel}

In the case of a 3D channel, the angular relaxation is dominated by small-angle scattering 
$|\Delta\bp| \ll p_F$, and therefore all angular harmonics have nearly the same relaxation time
$\tau^{-1} \sim T^2/E_F$ \cite{Gurzhi95-PRB, eval3D}. The exceptions are the spherical harmonics with $l=0$ and $l=1$, which have
zero relaxation rates because of the particle-number and momentum conservation laws. We assume that the
channel is cylindrically symmetric and  linearize Eq. \eqref{kinetic} with respect to the voltage drop assuming $eV \ll T$ by 
a substitution \cite{Haug}
\be
 f(\bn, \eps, x) = f_0(\eps) + f_0\,(1-f_0)\,\psi(x,\bn),
 \label{f-ansatz}
\ee
where $x$ is the longitudinal coordinate and $\psi(x,\bn)$ describes the angular distribution of electrons.
As the relaxation of all angular harmonics with $l>1$ may be approximately described by a single 
characteristic time $\tau$,  one may subtract the harmonics with $l<2$ from $\psi$ in the collision integral 
and write down Eq. \eqref{kinetic} for $\psi$  in the form
\begin{equation}
 v_x\,\dfrac{\partial\psi}{\partial x} =
 -\frac{1}{\tau}\,(\psi - \bar\psi - \psi_1),
 \label{psi3D-eq1}
\end{equation}
where $\bar\psi$ and $\psi_1$ are the zero and first harmonics of $\psi$ given by   
the angular integrals
\begin{gather}
 \bar\psi(x) = \int \frac{d\Omega}{4\pi}\,\psi(x, \theta), \label{harm_0-3D}
 \\
 \psi_1(x, \theta) = 3\cos\theta \int\frac{d\Omega'}{4\pi}\,\cos\theta'\,\psi(x,\theta'),
 \label{harm_1-3D}
\end{gather}
$\Omega$ is the solid angle in the momentum space, and $\theta$ is the angle between the momentum direction and the 
channel axis $x$.
Equation \eqref{psi3D-eq1} should be supplemented by the boundary conditions
\begin{gather}
 \psi(0,\, n_x>0) =  \frac{eV}{2T},
 \quad
 \psi(L,\, n_x<0) = -\frac{eV}{2T}.
 \label{boundary}
\end{gather}
Our goal is to obtain a closed set of equations for $\bar\psi$ and $\psi_1$. To this end, 
we first express $\psi(x,\theta)$ in terms of these quantities by means of Eq. \eqref{psi3D-eq1} and then again 
substitute it into Eqs. \eqref{harm_0-3D} and \eqref{harm_1-3D} to obtain self-consistency equations for them. 
The solution of \eqref{psi3D-eq1} can be obtained separately for right-moving $(\theta<\pi/2)$ and 
left-moving $(\theta>\pi/2)$ electrons by integrating its right-hand part along the trajectory emerging either 
from the left or right end of the channel \cite{Kulik77}. Hence 
\begin{multline}
 \psi(x, \theta) = \Theta(\pi/2-\theta)\,\psi_R(x,\theta) 
\\+ \Theta(\theta-\pi/2)\,\psi_L(x,\theta),
 \label{3D-R+L}
\end{multline}
where the right-moving and left-moving components are given by
\begin{subequations}\label{RL-sols}
\begin{align}
 \psi_R(x,\theta) = \frac{eV}{2T}\,e^{-t_R/\tau} 
 +\frac{1}{\tau} \int_0^{t_R} dt_R'\,  e^{-(t_R-t_R')/\tau}\,
\nonumber\\ \times
 [\bar\psi(t_R') + \psi_1(t_R',\theta)],
\label{psi_R-3D}\\
 \psi_L(x,\theta) = -\frac{eV}{2T}\,e^{-t_L/\tau} 
 +\frac{1}{\tau} \int_0^{t_L} dt_L'\,e^{-(t_L-t_L')/\tau}\,
\nonumber\\ \times
 [\bar\psi(t_L') + \psi_1(t_L',\theta)],
 \label{psi_L-3D}
\end{align}
\end{subequations}
and $t_R=x/(v_F\cos\theta)$,  $t_L = (L-x)/|v_F\cos\theta|$ are the traveling times of an electron from the left  
or right end of the channel to point $x$, respectively.  Now present the first harmonic of $\psi$ in the form
$\psi_1(x,\theta) = C \cos\theta$, where $C$ is independent of $x$ because of the current conservation. On
substitution of Eqs. \eqref{psi_R-3D} and \eqref{psi_L-3D} into 
Eq.~\eqref{harm_1-3D} one obtains a self-consistency equation 
\begin{multline}
  \left[ E_4\!\left(\frac{x}{l_{ee}}\right) + E_4\!\left(\frac{L-x}{l_{ee}}\right) \right] C
\\
 -\frac{1}{2}\,\frac{eV}{T}
  \left[ E_3\!\left(\frac{x}{l_{ee}}\right) + E_3\!\left(\frac{L-x}{l_{ee}}\right) \right]
\\
 =\int_0^L \frac{dx'}{l_{ee}}\,\sgn(x-x')\,E_2\!\left(\frac{|x-x'|}{l_{ee}}\right) \bar\psi(x'),
 \label{C-int3D}
\end{multline}
where $l_{ee} = v_F\tau$ and the quantities $E_n(x) = x^{n-1}\,\Gamma(1-n,x)$ are expressed in terms of
the incomplete gamma function. A similar self-consistency equation  may be obtained for $\bar\psi(x)$
(see \ref{perturbative}, Eq. \eqref{psi-int3D}), but it appears to be the result of differentiation of
Eq. \eqref{C-int3D} with respect to $x$ provided that $C$ is constant, hence there is only one independent 
equation for determining both $C$ and $\bar\psi(x)$. However Eq. \eqref{C-int3D} is a Fredholm equation of 
the first kind in $\bar\psi(x)$, which has a solution only if the left-hand side meets certain conditions.
Therefore there is no discretion in determining $C$ and $\bar\psi$.

As the first step, we solve the problem perturbatively. If $l_{ee} \to\infty$, it is easily seen that 
$\bar\psi^{(0)}(x)=0$.
Then one immediately obtains from Eq. \eqref{C-int3D} that $C^{(0)} = (3/4)\,eV/T$ and  arrives
at the standard expression for the Sharvin conductance \cite{Sharvin}
\be
 G^{(0)}_3 = \frac{e^2 S_0 p_F^2}{(2\pi)^2},
 \label{G03}
\ee
where $S_0$ is the cross-section of the channel. The first-order correction
in $L/l_{ee}$ to the conductance is of the form (see \ref{perturbative})
\be
 G_3^{(1)} = -\frac{1}{4}\,\frac{L}{l_{ee}}\,G^{(0)}_3.
 \label{G3-1}
\ee
Up to a numerical constant, this is the same result as in Ref. \cite{Nagaev12}.

\begin{figure}[t]
 \includegraphics[width=8.5cm]{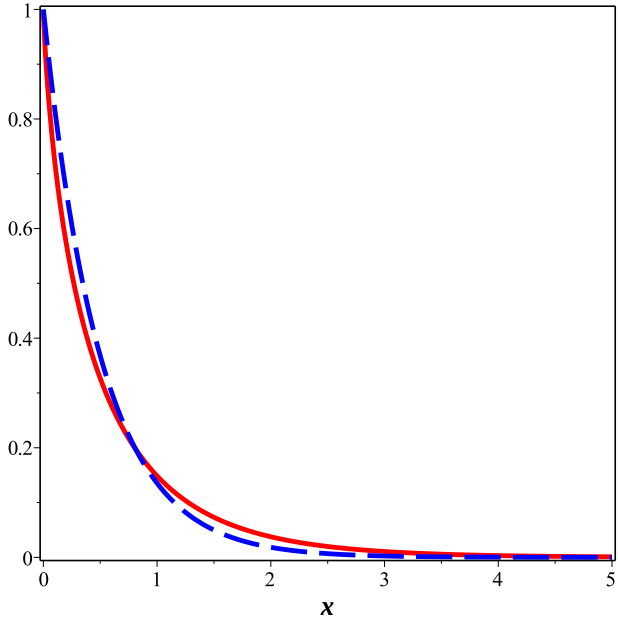}
 \caption{\label{fig:E2} Comparison of functions $E_2(x)$ (red solid line) and $\exp(-2x)$ (blue dashed line).}
\end{figure}

\begin{figure}
\includegraphics[width=8.5cm]{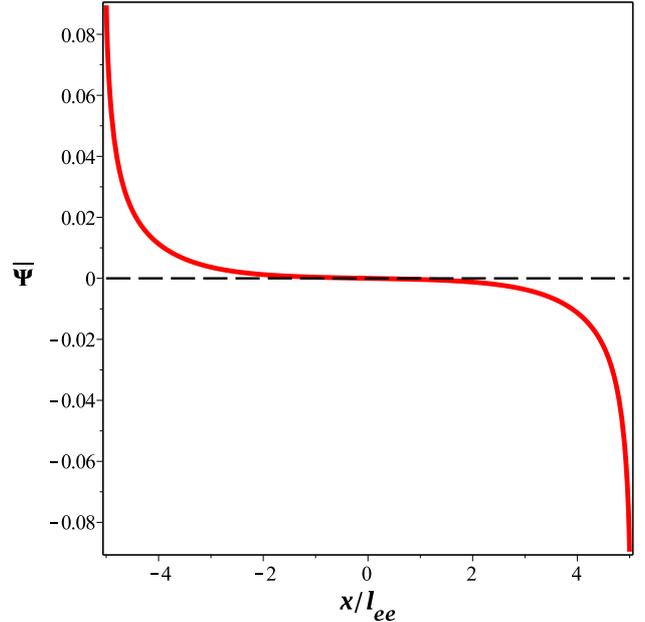}
\caption{\label{fig:psi} The isotropic part of the electron distribution $\bar\psi$ in units of
 $eV/T$ as a function of coordinate for $L/l_{ee}=10$.}
\end{figure}

If $L/l_{ee}$ is not small, Eq. \eqref{C-int3D} cannot be solved analytically. Because
this is a Fredholm equation of the first kind, its numerical solution is unstable with respect to rapid oscillations  
and cannot be obtained by the  standard methods \cite{Tikhonov77}. Therefore 
we use a semi-analytical approach and replace $E_2(x)$  in the kernel of Eq. \eqref{C-int3D} by $\exp(-2x)$.
This exponent coincides with $E_2(x)$  at $x=0$, is very close to it at $x\sim 1$ (see Fig.~\ref{fig:E2}), and bounds 
the same area from above.  The difference between these functions becomes significant only at $x\gg 1$, where both of them are exponentially small.  Unlike $E_2(x)$, this 
exponential allows an analytical solution of  Eq. \eqref{C-int3D} for 
arbitrary strength of  electron-electron scattering (see \ref{sec:int-sol-3D} for details). 
 This readily gives us the conductance of the channel in the form
\be
 G_3 = \frac{2}{3}\,G^{(0)}_3\,
 \frac{ 7 + 6\,E_3(L/l_{ee}) - 12\,E_4(L/l_{ee}) }{ 5 + 6\,E_4(L/l_{ee}) - 12\,E_5(L/l_{ee}) }.
 \label{G3-full}
\ee
Its weak-scattering expansion 
coincides to the first order with Eq. \eqref{G3-1}, and in the opposite limit $L/l_{ee}\to\infty$,
it tends to $(14/15)\,G^{(0)}_3\approx 0.93\,G^{(0)}_3$. The corresponding solution for $\bar\psi(x)$ is given by
\begin{multline}
 \bar{\psi}(x) = 
 \frac{1}{2}\Biggl[
  E_3\!\left(\frac{L-x}{l_{ee}}\right) -  E_3\!\left(\frac{x}{l_{ee}}\right)
  + 
  4\,E_5\!\left(\frac{x}{l_{ee}}\right) 
\\
  - 4\,E_5\!\left(\frac{L-x}{l_{ee}}\right)
 \Biggr] C
 -
 \frac{1}{4}\,\frac{eV}{T} \Biggl[
  E_2\!\left(\frac{L-x}{l_{ee}}\right) 
\\
  - E_2\!\left(\frac{x}{l_{ee}}\right)
 +
  4\,E_4\!\left(\frac{x}{l_{ee}}\right) - 4\,E_4\!\left(\frac{L-x}{l_{ee}}\right)
 \Biggr].
 \label{psi-full-3D}
\end{multline}
The coordinate dependence of $\bar\psi$ for $L/l_{ee}=10$ is shown in Fig. \ref{fig:psi}. It is almost zero
in the middle portion of the contact and sharply increases near its ends, so that its derivative
has a logarithmic singularity at $x=0$ and $x=L$. In the limit of strong scattering, the values of 
$\bar\psi$ at the ends of the channel tend to $\pm(11/120)\,eV/T$, which is well below its values in the 
reservoirs. The discontinuity of $\bar\psi$ at the ends of the channel is smeared if its finite width
is taken into account.

\section{2D channel}

Contrary to 3D systems, collisions
of electrons with almost opposite momenta play an essential role in the angular relaxation of electron distribution 
in a 2D system \cite{Gurzhi95-PRB,Ledwith17}. This results in a sharp difference in the relaxation  of 
symmetric and antisymmetric parts of the distribution function in the momentum space. As this type of 
scattering just rotates a pair of electrons with opposite momenta in the $\bp$ space about $\bp=0$, it 
affects the symmetric part of electron distribution but does not affect the antisymmetric one \cite{Gurzhi95-PRB}.
 As a result, 
the relaxation rate for the symmetric part $\tau_s^{-1} \sim T^2/E_F$ is parametrically larger than
the relaxation rate for the antisymmetric part $\tau_a^{-1} \sim T^4/E_F^3$. 
Therefore we separate the collision integral into the symmetric and antisymmetric parts and describe each of them
by its own relaxation time. 
Though the relaxation rate grows faster with harmonic index for odd harmonics than for even ones, this approximation
is sufficient for determining the parametric dependence of the correction to the conductance because it is dominated 
by harmonics with indices much smaller than $E_F/T$. 
In view of this, one may write down the kinetic equation for $\psi$ in the 
form
\be
 v_x\,\frac{\partial\psi}{\partial x} = -\frac{1}{\tau_s}\,(\psi_s - \bar\psi)
 -\frac{1}{\tau_a}\,(\psi_a - \psi_1),
 \label{psi2D-eq1}
\ee
where $\psi_{s,a} = [\psi(\bn) \pm \psi(-\bn)]/2$ are the symmetric and antisymmetric parts of $\psi$. 
The zero and first harmonics of $\psi$ are defined as
\begin{gather}
 \bar\psi(x) = \frac{1}{\pi} \int_0^{\pi} d\p\,\psi(x,\p),
 \label{harm_0-2D}\\
 \psi_1(\p) = \frac{2}{\pi}\,\cos\p \int_0^{\pi} d\p'\,\cos\p'\,\psi(x,\p') \equiv C\cos\p,
 \label{harm_1-2D}
\end{gather}
where $\p$ is the angle between $\bn$ and the longitudinal axis $x$ of the channel. As in the 3D case, they are
not affected by the collisions.

Our goal is to express $\psi(x,\p)$ in terms of $\bar\psi$ and $\psi_1$ and then to obtain for them 
self-consistency equations by means of Eqs. \eqref{harm_0-2D} and \eqref{harm_1-2D}, much like in
the 3D case. As the first step, we form
symmetric and antisymmetric combinations of Eq. \eqref{psi2D-eq1}  for $\bn$ and $-\bn$ to obtain a 
system of equations for $\psi_s$ and $\psi_a$ in the form
\begin{subequations}\label{s,a-eqs}
\begin{align}
 &|v_x|\,\frac{d\psi_a}{dx} = -\frac{1}{\tau_s}\,(\psi_s - \bar\psi),
 \label{a-eq}\\
 &|v_x|\,\frac{d\psi_s}{dx} = -\frac{1}{\tau_a}\,(\psi_a - \psi_1).
 \label{s-eq}
\end{align}
\end{subequations}
To diagonalize this system, we introduce new variables
$ \psi_{\pm} = \sqrt{\tau_a}\,\psi_s \pm \sqrt{\tau_s}\,\psi_a$
and bring it to the form
\begin{subequations}\label{+-eqs}
\begin{align}
 &|v_x|\,\frac{d\psi_{+}}{dx} + \frac{1}{\tau_m}\,\psi_{+}
 = \frac{1}{\sqrt{\tau_s}}\,\bar\psi + \frac{1}{\sqrt{\tau_a}}\,\psi_1
 \label{+eq}\\
 &|v_x|\,\frac{d\psi_{-}}{dx} - \frac{1}{\tau_m}\,\psi_{-}
 = \frac{1}{\sqrt{\tau_s}}\,\bar\psi - \frac{1}{\sqrt{\tau_a}}\,\psi_1,
 \label{-eq}
\end{align}
\end{subequations}
where $\tau_m = \sqrt{\tau_s \tau_a}$ is a new characteristic relaxation time that takes into account
the coupling of odd and even angular harmonics by the gradient terms. Making use of the inverse
transform $\psi_a=\frac{1}{2}(\psi_{+} + \psi_{-})/\sqrt{\tau_s}$ and $\psi_s=\frac{1}{2}(\psi_{+}-\psi_{-})/\sqrt{\tau_a}$,
one may express $\psi(\p)$ in terms of these functions as
\be
 \psi = 
 \begin{cases} 
  \psi_s + \psi_a = s_{+}\psi_{+} + s_{-}\psi_{-}, & \p<\pi/2
  \\
  \psi_s - \psi_a = -s_{-}\psi_{+} - s_{+}\psi_{-}, & \p>\pi/2
 \end{cases}
 \label{psi_vs_+-}
\ee
where $s_{\pm} = \frac{1}{2}(\sqrt{\tau_a} \pm \sqrt{\tau_s})/\tau_m$.

Similarly to Eqs. \eqref{RL-sols}, 
the solutions of Eqs. \eqref{+-eqs} may be written as integrals of their right-hand parts along the trajectories
emerging from the left and right ends of the channel
\begin{subequations}\label{+-sols}
\begin{align}
 \psi_{+}(x) = \psi_{+}(0)\,e^{-t_R/\tau_m}
 +\int_0^{t_R} dt_R'\,e^{-(t_R-t_R')/\tau_m}\,
\nonumber\\ \times
 \left[\frac{1}{\sqrt{\tau_s}}\,\bar\psi(t_R') + \frac{1}{\sqrt{\tau_a}}\,\psi_1(t_R')\right],
 \label{+sol}\\
 \psi_{-}(x) = \psi_{-}(L)\,e^{-t_L/\tau_m}
 -\int_0^{t_L} dt_L\,e^{-(t_L-t_L')/\tau_m}\,
\nonumber\\ \times
 \left[\frac{1}{\sqrt{\tau_s}}\,\bar\psi(t_L') - \frac{1}{\sqrt{\tau_a}}\,\psi_1(t_L')\right],
 \label{-sol}
\end{align}
\end{subequations}
where $t_R = x/|v_x|$ and $t_L = (L-x)/|v_x|$ are defined as in Eqs. \eqref{RL-sols} and the angular argument $\p$ is 
omitted for brevity. However
in contrast to Eqs. \eqref{RL-sols}, the initial conditions for Eqs. \eqref{+-eqs} $\psi_{+}(0,\p)$ and $\psi_{-}(L,\p)$ are now unknown quantities themselves, 
as well as $\psi_{+}(L,\p)$ and $\psi_{-}(0,\p)$. To also express these four values in terms of $\bar\psi$ and $\psi_1$, 
one needs four equations. Two of
them may be obtained by substituting $x=L$ into Eq. \eqref{+sol} and $x=0$ into Eq. \eqref{-sol}. Another pair
of equations may be obtained from the boundary conditions Eqs. \eqref{boundary} and \eqref{psi_vs_+-}. It reads
\begin{multline}\label{boundary+-}
 s_{+}\psi_{+}(0,\p) + s_{-}\psi_{-}(0,\p)
\\ = 
 s_{-}\psi_{+}(L,\p) + s_{+}\psi_{-}(L,\p)
 = eV/2T.
\end{multline}
The solutions of this system are substituted into Eqs. \eqref{+-sols} and the resulting
$\psi_{+}(x,\p)$ and $\psi_{-}(x,\p)$ are used to express $\psi(x,\p)$
in terms of $\bar\psi(x)$ and $C$ by means of \eqref{psi_vs_+-}. Upon the substitution of $\psi(x,\p)$ into Eqs. \eqref{harm_0-2D}
and \eqref{harm_1-2D} one obtains self-consistency equations for these quantities.
In the general case, these  equations are too 
cumbersome to be presented here. In the limit of a long channel $L \gg l_m \equiv v_F\tau_m$, the 
self-consistency equation for $C$ is of the form
\begin{multline}
 \sqrt{\frac{\tau_a}{\tau_s}}\,\int\limits_0^L \frac{dx'}{l_m}\,
 \Biggl[ \kappa_0\,\tilde{E}_2\!\left(\frac{x+x'   }{l_m}\right)
        -\kappa_0\,\tilde{E}_2\!\left(\frac{2L-x-x'}{l_m}\right)
\\
        +\sgn(x-x')\,
                   \tilde{E}_2\!\left(\frac{|x-x'| }{l_m}\right)
 \Biggr]\,\bar\psi(x')
\\
 -(1 + \kappa_0)\,C\,
 \Biggl[ \tilde{E}_4\!\left(\frac{x  }{l_m}\right)
        +\tilde{E}_4\!\left(\frac{L-x}{l_m}\right)
 \Biggr]
\\
 +\frac{eV}{T}\,\sigma_0\,
 \Biggl[ \tilde{E}_3\!\left(\frac{x  }{l_m}\right)
        +\tilde{E}_3\!\left(\frac{L-x}{l_m}\right)
 \Biggr] = 0,
 \label{C-int2D}
\end{multline}
where 
\be
 \kappa_0 = \frac{\tau_a - \tau_s}{(\sqrt{\tau_a} + \sqrt{\tau_s})^2},
 \quad
 \sigma_0 = \frac{\sqrt{\tau_a}}{\sqrt{\tau_a}+\sqrt{\tau_s}},
\ee
and
\be
 \tilde{E}_n(x) = \int_1^{\infty} d\xi\, \frac{e^{-\xi x}}{\xi^{n-1}\,\sqrt{\xi^2-1}}.
\ee
As in the 3D case, the self-consistency equation for $\bar\psi(x)$ may be obtained by differentiating 
Eq. \eqref{C-int2D} with respect to $x$ and therefore gives no additional information. Much like
Eq. \eqref{C-int3D}, it is a Fredholm equation of the first kind and determines uniquely both $C$ and
$\bar\psi(x)$. Similarly to the 3D case, Eq. \eqref{C-int2D} may be approximately solved by replacing 
$\tilde{E}_2(x)$ in the integrand with $(\pi/2)\exp(-\pi x/2)$. The replacement function is chosen such
that it coincides with $\tilde{E}_2(x)$ at $x=0$ and bounds the same area from above. 
With this replacement, one can analytically calculate $C$ (see \ref{sec:int-sol-2D}) and 
obtain the conductance in the form
\be
 G_2
 = \frac{3}{8}\,\frac{8 + \pi^2\,\sqrt{\tau_s/\tau_a}}{3 + 4\,\sqrt{\tau_s/\tau_a}}\,{G_{02}},
 \label{G2}
\ee
where $G_{02} = e^2 p_F W/\pi^2$ is the Sharvin conductance of a 2D ballistic contact and $W$ is the 
width of the channel. This suggests that $G_2$ is a monotonically decreasing function of $\tau_s/\tau_a$.
In the limit $\tau_s\ll\tau_a$, it approximately equals
\be
 G_2 \approx  \left(1 - 0.1\,\sqrt{\frac{\tau_s}{\tau_a}}\right) {G_{02}}.
 \label{G2-asympt}
\ee
Hence in the limit of a long 2D channel, the negative correction to the conductance saturates at a value
proportional to the temperature.

\section{Discussion}

The inelastic correction to the current is proportional to the total rate of collisions in the channel that
change the number of left-moving and right-moving electrons. If the channel is short, the
distribution function of electrons is almost constant inside it, and this rate  
is proportional to its length and the relaxation rate of the momentum-antisymmetric part of the electron distribution. 
However if the channel is much longer than a certain relaxation length, the electron distribution 
becomes strongly coordi\-nate-dependent. It sharply changes in space and contains a large number of angular harmonics
near the ends of the channel, but  the electrons in its
middle part are described by an almost coordinate-independent quasi-equilibrium Fe\-rmi distribution with a shifted 
center of mass that accounts
for the current flow. This distribution identically turns the collision integral into zero, and therefore
only the scattering near the ends of the channel affects the current. Hence the correction to the current 
saturates in the limit of a long channel and is proportional to the product of the antisymmetric relaxation rate
and the relaxation length of the electron distribution. 
In the case of a 3D channel, both the even and odd angular harmonics relax at the same rate $\tau^{-1}$ while 
the relaxation length is proportional to $\tau$. Therefore the product of these quantities is constant, and in the 
limit of a long channel, the correction to the conductance is about 7\% regardless of the scattering strength.

In the case of a 2D channel, there are two different relaxation rates for the antisymmetric and symmetric parts
of the electron distribution, $\tau_a^{-1}$ and $\tau_s^{-1}$. However the spatial relaxation of these parts
to the quasi-equilibrium Fermi distribution in the middle part of the channel is not independent. Because the gradient
term in the kinetic equation mixes these parts together, the resulting relaxation length is proportional  to
$\tau_m = \sqrt{\tau_a\tau_s}$ for both of them. Therefore the resulting relative correction to the conductance 
$(G_2-G_{02})/G_{02}$ is 
proportional to $\tau_a^{-1}\tau_m = \sqrt{\tau_s/\tau_a} \propto T$.

A saturation of the correction to the conductance was predicted in Ref. \cite{Rech09} 
for a long 
single-mode quantum wire where it resulted from three-electron collisions. The authors obtained that the 
correction  is determined only by conservation laws and does not depend on the details of 
scattering, but this is not the case for a semiclassical system.

An experimental verification of Eqs. \eqref{G3-full} and \eqref{G2} would be a good test of the Gurzhi
theory of electron-electron relaxation in a 2D gas. In experiments on AlGaAs/GaAs 
heterostructures \cite{Renard08,Melnikov12}, the elastic mean free path due to impurity scattering was about 
20~$\mu$m, $E_F$ was 2.9 meV, and $v_F$ was $1.3\times 10^7$ cm/s. Together with the estimate of the interaction
parameter \cite{Melnikov12} $1 < \alpha_{ee} <2$, this suggests that $l_{ee} \approx \alpha_{ee} \hbar v_F/(k_BT)^2$
will be smaller already at $T\ge 1.5$ K. The strength of boundary scattering is hard to estimate, but there are 
indications \cite{Molenkamp94} that in the case of a channel formed by remote electrostatic gates, 80\% of all 
boundary collisions are specular. Probably their percentage may be increased further by increasing the distance 
between the channel and the gates.
Therefore the regime discussed above is experimentally attainable. The predicted
saturation of the correction to the resistance may be observed, e. g., by increasing the temperature at a fixed
length of the channel. 
Though there is some uncertainty in the numerical prefactor of the correction to the conductance in the 2D case,
it can be distinguished by its linear temperature dependence.

\section{Summary}

In summary, we have calculated the correction to the conductance of a long many-mode ballistic channel
that results from electron-electron scattering. In the case of a sufficiently long 3D channel, the resulting correction
is independent of temperature and the parameter of electron-electron scattering because the rate of collisions
affecting the current is comparable with total collision rate that forms the shape of the electron distribution 
function. In the case of a 2D channel, the rate of collisions affecting the current is much smaller than 
the total collision rate, and the resulting saturation value of negative correction to the conductance is proportional to the temperature. The characteristic length of channel that corresponds to the saturation  
in the 2D case is different from the standard electron--electron scattering length.

\section*{Acknowledgments}
This work was supported by Russian Foundation for Basic Research, grant 16-02--00583-a.

\appendix

\section{Perturbative calculation of the correction to the current in 3D case}
\label{perturbative}

 The first-order correction
to $\bar\psi$ in electron-electron scattering  may be obtained by expanding $E_n(x/l_{ee})$  in $x/l_{ee}$
and substituting $\bar\psi^{(0)}$ and $C^{(0)}$ into the self-consistency equation for $\bar\psi(x)$

\begin{multline}
 \bar\psi(x) = \frac{1}{4}\,\frac{eV}{T}
               \left[ E_2\!\left(\frac{x}{l_{ee}}\right) - E_2\!\left(\frac{L-x}{l_{ee}}\right) \right]
\\
             + \frac{1}{2}
               \left[ E_3\!\left(\frac{L-x}{l_{ee}}\right) - E_3\!\left(\frac{x}{l_{ee}}\right) \right] C
\\
             + \frac{1}{2}\int_0^L \frac{dx'}{l_{ee}}\,E_1\!\left(\frac{|x-x'|}{l_{ee}}\right) \bar\psi(x'),
 \label{psi-int3D}
\end{multline}
which gives
\begin{multline}
 \bar\psi^{(1)}(x) = \frac{1}{4}\,\frac{eV}{T}
 \Biggl\{
  \frac{x}{l_{ee}} \left[ \frac{1}{2} + \gamma + \ln\!\left(\frac{x}{l_{ee}}\right)\right]
\\
  -
  \frac{L-x}{l_{ee}} \left[ \frac{1}{2} + \gamma + \ln\!\left(\frac{L-x}{l_{ee}}\right)\right]
 \Biggr\},
 \label{psi-1-3D}
\end{multline}
where $\gamma=0.577$ is the Euler constant. A substitution of these quantities into Eq. \eqref{C-int3D} 
results in a correction to $C$ and hence to the conductance Eq. \eqref{G3-1}.

\section{Solution of the integral equation for the 3D case}
\label{sec:int-sol-3D}

Consider the integral equation
\be
 \int_{0}^{L} dx'\,\sgn(x-x')\,e^{-\lambda|x-x'|}\,\bar\psi(x') = g(x)
 \label{F1a}
\ee
with an antisymmetric kernel. Our goal is to determine the conditions on which it has a solution. 
To this end, we consider an auxiliary equation
\be
 \int_{0}^{L} dx'\,e^{-\lambda|x-x'|}\,\bar\psi(x') = g_1(x)
 \label{F1s}
\ee
with a symmetric kernel. The differentiation of this equation with respect to $x$ gives Eq. (\ref{F1a}) 
provided that
\be
 g(x) = -\frac{1}{\lambda}\,\frac{dg_1}{dx}.
 \label{g-vs-g1}
\ee
Differentiating Eq. \eqref{F1s} with respect to $x$ for the second time gives a Fredholm equation of the 
second type
\be
 \bar\psi(x) - \frac{\lambda}{2} \int\limits_{0}^{L} dx'\,e^{-\lambda|x-x'|}\,\bar\psi(x')
 = 
 -\frac{1}{2\lambda}\,\frac{d^2 g_1}{dx^2}.
 \label{F2}
\ee
The integral in left-hand side of Eq. (\ref{F2}) may be excluded by means of Eq. (\ref{F1s}), so one
obtains
\be
 \bar\psi(x) = \frac{1}{2\lambda}
 \left[ \lambda^2 g_1(x) - \frac{d^2 g_1}{dx^2} \right].
 \label{psi-gen}
\ee
Now we have to make sure that $\bar\psi(x)$ from Eq. (\ref{psi-gen}) also satisfies Eq. (\ref{F1s}). To this end,
we substitute it into the left-hand side of Eq. (\ref{F1s}) and integrate twice by parts. Thus it is brought
to the form
\begin{multline}
 g_1(x) 
 - \frac{1}{2\lambda}\,e^{-\lambda(L-x)}
 \left.\left( \frac{dg_1}{dx} + \lambda g_1 \right)\right|_{x=L}
\\
 + \frac{1}{2\lambda}\,e^{-\lambda x}
  \left.\left( \frac{dg_1}{dx} - \lambda g_1 \right)\right|_{x=0},
\end{multline}
hence the solution of Eq. (\ref{F1s}) exists and is given by (\ref{psi-gen}) if
\be
 \left.\left( \frac{dg_1}{dx} + \lambda g_1 \right)\right|_{x=L}
 =
 \left.\left( \frac{dg_1}{dx} - \lambda g_1 \right)\right|_{x=0}
 =0.
 \label{cond-1}
\ee
Rewrite now this condition in terms of $g(x)$ by means of (\ref{g-vs-g1}). If $g(x-L/2)$ is an even function
of $x$, $g_1(x-L/2)$ must be an odd function of $x$. 
\be
 g_1(x) = -\lambda \int_{L/2}^x dx'\,g(x').
 \label{g1}
\ee
The condition (\ref{cond-1}) at $x=L$ takes up the form
\be
 g(L) + \lambda \int_{L/2}^{L} dx'\,g(x') = 0.
 \label{F1-cond3D}
\ee
If it is satisfied, the condition (\ref{cond-1}) at $x=-L/2$ is also met because $\tilde{f}$ is an odd function.
Correspondingly,
\be
 \bar\psi(x) = \frac{1}{2}\,\frac{dg}{dx} - \frac{\lambda^2}{2} \int_{L/2}^x dx'\,g(x').
 \label{F1-sol3D}
\ee
 One easily obtains a 
linear equation for $C$ by substituting
\begin{multline}
 g(x) = 
 \left[ E_4\!\left(\frac{x}{l_{ee}}\right) + E_4\!\left(\frac{L-x}{l_{ee}}\right) \right] C
\\
 -\frac{1}{2}\,\frac{eV}{T}
 \left[ E_3\!\left(\frac{x}{l_{ee}}\right) + E_3\!\left(\frac{L-x}{l_{ee}}\right) \right]
 \label{g-3D}
\end{multline}
into Eq. \eqref{F1-cond3D}.

\section{Solution of the integral equation for the 2D case}
\label{sec:int-sol-2D}

The integral equation for $\bar\psi(x)$ in the 2D case may be written in the form
\begin{multline}
 \int_0^L dx'\,\Biggl[
   \kappa_0\,e^{-\lambda\,(x+x')} 
  -\kappa_0\,e^{\lambda\,(x+x'-2L)}
\\
  +\sgn(x-x')\,e^{-\lambda\,|x-x'|}
 \Biggr]\,\bar\psi(x') 
 = \tilde{g}(x).
 \label{F1a-2D}
\end{multline}
To find the solution of Eq. \eqref{F1a-2D} and the condition for its existence, we consider
an auxiliary equation
\begin{multline}
 \int_0^L dx'\,\Biggl[
   \kappa_0\,e^{-\lambda\,(x+x')} 
  +\kappa_0\,e^{\lambda\,(x+x'-2L)}
\\
  + e^{-\lambda\,|x-x'|}
 \Biggr]\,\bar\psi(x') 
 = \tilde{g}_1(x).
 \label{F1s-2D}
\end{multline}
The differentiation of both sides of this equation with respect to $x$ gives precisely Eq. \eqref{F1a-2D} 
provided that
\be
 \tilde{g}(x) = -\frac{1}{\lambda}\,\frac{d\tilde{g}_1}{dx}.
 \label{g1-2D}
\ee
By differentiating Eq. \eqref{F1s-2D} twice with respect to $x$, one obtains
\begin{multline}
 \lambda^2\, \int_0^L dx'\,\Biggl[
   \kappa_0\,e^{-\lambda\,(x+x')} 
  +\kappa_0\,e^{\lambda\,(x+x'-2L)}
\\
  + e^{-\lambda\,|x-x'|}
 \Biggr]\,\bar\psi(x') 
 -2\lambda\,\bar\psi(x) = \frac{d^2\tilde{g}_1}{dx^2}.
 \label{F2-2D}
\end{multline}
In view of Eq. \eqref{F1s-2D}, it may be recast in the form
\be
 \lambda^2\,\tilde{g}_1(x) - 2\,\lambda\,\bar\psi(x) = \frac{d^2\tilde{g}_1}{dx^2},
 \label{psi-eq2D}
\ee
hence
\be
 \bar\psi(x) = \frac{1}{2\lambda} \left[\lambda^2\,\tilde{g}_1(x) - \frac{d^2\tilde{g}_1}{dx^2}\right].
 \label{psi-sol1-2D}
\ee
So if Eq. \eqref{F1s-2D} has a solution, it is of the form \eqref{psi-sol1-2D}. Substitute now
Eq. \eqref{psi-sol1-2D} back into Eq. \eqref{F1s-2D} and check whether it is satisfied.
To do this, 
we perform twice the integration by parts in its left-hand side to get rid of the derivatives with respect to
$x$. Upon these integrations, the left-hand side of Eq. \eqref{F1s-2D} assumes the form
\begin{multline}
 \int_0^L dx'\,\Biggl[\kappa_0\,e^{-\lambda\,(x+x')}+\kappa_0\,e^{\lambda\,(x+x'-2L)}
 + e^{-\lambda\,|x-x'|} \Biggr] 
\\ \times
 \frac{1}{2\lambda} \left[\lambda^2\,\tilde{g}_1(x) - \frac{d^2\tilde{g}_1}{dx^2}\right]
\\
 = \tilde{g}_1(x)
 -\frac{1}{2}\,e^{\lambda\,(x-L)}\,
 \left[(1-\kappa_0)\,\tilde{g}_1 + (1+\kappa_0)\,\frac{1}{\lambda}\,\frac{d\tilde{g}_1}{dx}\right]_{x=L}
\\
 -\frac{1}{2}\,e^{-\lambda\,x}\,
 \left[(1-\kappa_0)\,\tilde{g}_1 - (1+\kappa_0)\,\frac{1}{\lambda}\,\frac{d\tilde{g}_1}{dx}\right]_{x=0}
\\
 -\frac{1}{2}\,e^{-\lambda\,(x+L)}\,\kappa_0
 \left(\tilde{g}_1 + \frac{1}{\lambda}\,\frac{d\tilde{g}_1}{dx}\right)_{x=L}
\\
 -\frac{1}{2}\,e^{\lambda\,(x-2L)}\,
 \left(\tilde{g}_1 - \frac{1}{\lambda}\,\frac{d\tilde{g}_1}{dx}\right)_{x=0}.
 \label{by-parts-2D}
\end{multline}
The two last terms in this equation are exponentially small and may be omitted. Hence the solution
of Eq. \eqref{F1s-2D} exists only if 
\begin{multline}
 (1-\kappa_0)\,\tilde{g}_1 + (1+\kappa_0)\,\frac{1}{\lambda}\,\left.\frac{d\tilde{g}_1}{dx}\right|_{x=L}
\\ =
 (1-\kappa_0)\,\tilde{g}_1 - (1+\kappa_0)\,\frac{1}{\lambda}\,\left.\frac{d\tilde{g}_1}{dx}\right|_{x=0}
 =0.
 \label{F1-cond2D}
\end{multline}
Using the relation \eqref{g1-2D}, one obtains that 
\be 
 (1+\kappa_0)\,\tilde{g}(L) + \lambda\,(1-\kappa_0) \int_{L/2}^L dx\,\tilde{g}(x) =0.
 \label{g-cond-2D}
\ee
As  $\tilde{g}(x-L/2)$ is even function of $x$, this ensures the fulfilment of both
equations \eqref{F1-cond2D}.

A substitution of $\lambda=\pi/2l_m$ and
\begin{multline}
 \tilde{g}(x) = (1+\kappa_0)\,C
 \Biggl[ \tilde{E}_4\!\left(\frac{x  }{l_m}\right)
        +\tilde{E}_4\!\left(\frac{L-x}{l_m}\right)
 \Biggr]
\\
 -\frac{eV}{T}\,\sigma_0\,
 \Biggl[ \tilde{E}_3\!\left(\frac{x  }{l_m}\right)
        +\tilde{E}_3\!\left(\frac{L-x}{l_m}\right)
 \Biggr]
 \label{g-2D}
\end{multline}
into Eq. \eqref{g-cond-2D} readily gives  $C$ and Eq. \eqref{G2}.


\begin{thebibliography}{99}
%
\bibitem{Peierls}
R. Peierls, Ann. Phys. (Leipzig) 395 (1929) 1055 .
%
\bibitem{Molenkamp94}
L. W. Molenkamp and M. J. M. de Jong, Phys. Rev. B  49 (1994) 5038.
%
\bibitem{Black80}
J. E. Black, Phys. Rev. B  21 (1980) 3279.
%
\bibitem{Gurzhi63}
R. N. Gurzhi, Pis'ma Zh. Eksp. Teor. Fiz.  44 (1963) 771  [JETP Lett.  17 (1963) 521];
Usp. Fiz. Nauk  94 (1968) 689  [Sov. Phys. Usp.  11 (1968) 255].
%
\bibitem{Guo17}
H. Guo, E. Ilseven, G. Falkovich, and L. S. Levitov, PNAS  114 (2017) 3068.
%
\bibitem{Nagaev08}
K. E. Nagaev and O. S. Ayvazyan,  Phys. Rev. Lett.  101 (2008) 216807.
%
\bibitem{Nagaev10}
K. E. Nagaev and T. V. Kostyuchenko, Phys. Rev. B  81 (2010) 125316.
%
\bibitem{Renard08}
V. T. Renard, O. A. Tkachenko, V. A. Tkachenko, T. Ota, N. Kumada, J. C. Portal, and Y. Hirayama, 
Phys. Rev. Lett.  100 (2008) 186801.
%
\bibitem{Melnikov12}
M. Yu. Melnikov, J. P. Kotthaus, V. Pellegrini, L. Sorba, G. Biasiol, and V. S. Khrapai, 
Phys. Rev. B  86 (2012) 075425.
%
\bibitem{Nagaev12}
K. E. Nagaev and N. Yu. Sergeeva, Phys. Rev. B { 85} (2012) 165404.
%
\bibitem{Lunde07} 
A. M. Lunde, K. Flensberg, and L. I. Glazman, Phys. Rev. B 75 (2007) 245418.
%
\bibitem{Gurzhi95-PRB}
R.N. Gurzhi, A.N. Kalinenko, and A.I. Kopeliovich, Phys. Rev. Lett.  74 (1995) 3872;
Phys. Rev. B  52 (1995) 4744.
%
\bibitem{Ledwith17}
P. Ledwith, H. Guo, and L. Levitov,  arXiv:1708.01915.
%
\bibitem{eval3D}
Exact calculations show that the relaxation rates for angular harmonics with $l>1$ differ by no more than 30\%,
to be published elsewhere.
\bibitem{Haug}
H. J. W.  Haug and A.-P. Jauho, Quantum Kinetics in Transport and Optics of Semiconductors, Springer-Verlag Berlin, 2008.
%
\bibitem{Kulik77}
I. O. Kulik, R. I. Shekhter, and A. N. Omelyanchouk, Solid State Comm.  23 (1977) 301;
I. O. Kulik, A. N. Omel'yanchuk, and R. I. Shekhter, Fiz. Nizk. Temp.  3 (1977) 1543   
[Sov. J. Low Temp. Phys.  3 (1977) 740].
%
\bibitem{Sharvin}
Y. V. Sharvin, Zh. Eksp. Teor. Fiz.  48 (1965) 984  [Sov. Phys. JETP  21 (1965) 655].
%
\bibitem{Tikhonov77}
A. Tikhonov  and V. Arsenin, Solutions of Ill-posed Problems, Winston and Sons, Washington, 1977.
%
\bibitem{Rech09} 
J. Rech, T. Micklitz, and K. A. Matveev, Phys. Rev. Lett.  102 (2009) 116402.
%

\end{thebibliography}
\end{document}